\documentclass{elsarticle}

\usepackage{graphicx}
\usepackage{lineno,hyperref}
\usepackage{epsfig,amsmath,amssymb,natbib}
\usepackage{color}
\modulolinenumbers[5]

\journal{}

\newcommand{\textin}[1]{\mbox{\scriptsize{#1}}}

\begin{document}

\begin{frontmatter}

\title{Viscoelastic Liquid Bridge Breakup and Liquid Transfer Between two Surfaces}

\author[label1]{H. Chen}
\author[label2]{A. Ponce-Torres}
\author[label2]{J. M. Montanero}
\author[label1]{A. Amirfazli\corref{cor1}}
\address[label1]{Department of Mechanical Engineering, York University, Toronto, ON, M3J 1P3, Canada}
\address[label2]{Departamento\ de Ingenier\'{\i}a Mec\'anica, Energ\'etica y de los Materiales and\\ 
Instituto de Computaci\'on Cient\'{\i}fica Avanzada (ICCAEx),\\ Universidad de Extremadura, E-06006 Badajoz, Spain}
\cortext[cor1]{Corresponding author: A. Amirfazli (alidad2@yorku.ca)}

\begin{abstract}
We studied experimentally the breakup of liquid bridges made of aqueous solutions of Poly(acrylic acid) between two separating solid surfaces with freely moving contact lines. For polymer concentrations higher than a certain threshold ($\sim$ 30 ppm), the contact line on the surface with the highest receding contact angle fully retracts before the liquid bridge capillary breakup takes place at its neck. This means that all the liquid remains attached to the opposing surface when the surfaces are separated. This behavior occurs regardless of the range of liquid volume and stretching speed studied. Such behavior is very different from that observed for Newtonian liquids or non-Newtonian systems where contact lines are intentionally pinned. It is shown that this behavior stems from the competition between thinning of bridge neck (delayed by extensional thickening) and receding of contact line (enhanced by shear thinning) on the surface with lower receding contact angle. If the two surfaces exhibit the same wetting properties, the upper contact line fully retracts before the capillary breakup due to the asymmetry caused by gravity, and, therefore, all the liquid remains on the lower surface.
\end{abstract}


\end{frontmatter}



\section{Introduction}
\label{sec1}

The liquid bridge has been studied for a long time both at the fundamental and practical levels \citep{MP20}. Liquid bridges formed between two solid surfaces are relevant to many applications such as electro-wetting-assisted drop deposition,  printing of micro-scale electronic circuits, semiconductors and biological micro-arrays, capillary gripping and packaging, and offset printing. In the latter case, ink is transferred by stretching an ink bridge formed between the donor and acceptor surfaces \citep{K15}. The volume of liquid transferred onto the acceptor surface is important because a small amount of ink remaining on the donor surface can produce image quality defects. Many inks of relevance contain particles and macromolecules that substantially change the liquid dynamical response. Such inks exhibit complex rheological behavior such as shear thinning and extensional thickening. The effects of these factors on the transfer between two surfaces is poorly understood.

Consider a Newtonian liquid bridge of density $\rho$, viscosity $\eta$, and surface tension $\gamma$ between two horizontal solid surfaces. The triple contact lines are allowed to freely move over those surfaces. The lower (donor) surface remains still while the upper (acceptor) one moves at a constant speed $v$ away from the lower surface. At the initial instant, the contact angles on acceptor and donor surfaces take values within the interval defined by the advancing and receding contact angles for the corresponding surface. As the liquid bridge stretches, contact angle on both surfaces decrease. When contact angles become smaller than the corresponding receding ones, the triple contact line moves \citep{SA13}. 

\begin{figure}
\begin{center}
\resizebox{0.65\hsize}{!}{\includegraphics{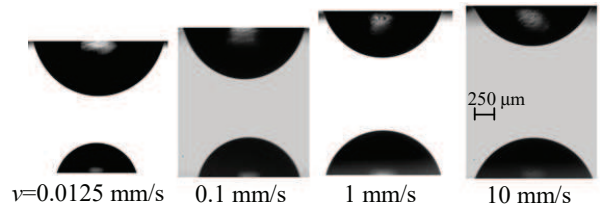}}
\end{center}
\caption{Transfer of glycerol between between (AL) (donor surface) and PEMA (acceptor surface) for different stretching velocities. The scale bar applies to all the images. The droplet volume was 1.2 $\mu$l.}
\label{glycerol}
\end{figure}

Three possible regimes can be identified depending on the value of the Capillary number $\text{Ca}=\eta v/\gamma$: (i) the quasi-static regime, which takes place for vanishing Ca and liquid transfer is controlled by the wettability (contact angles) of the surfaces; (ii) the dynamic regime, which arises for sufficiently large Ca and is dominated by viscous/inertial forces; and (iii) the transition regime, which occurs for intermediate values of Ca and results from the competition of wettability and viscous/inertial forces. The transfer ratio (the volume of liquid transferred to the acceptor surface over the total liquid volume) due to bridge breakup in the quasi-static regime, $\alpha_0$, can be calculated from the empirical formula \citep{CTA15a}:
\begin{equation}
\label{e1}
\alpha_0=\frac{1}{1+\exp[-3.142 (\theta_r^{(\textin{acc})}+\theta_r^{(\textin{don})})^{2.528} (\theta_r^{(\textin{don})}-\theta_r^{(\textin{acc})})]},
\end{equation}
where $\theta_r^{(\textin{acc})}$ and $\theta_r^{(\textin{don})}$ are the receding contact angles of the acceptor and donnor surfaces, respectively. Numerical simulations  are in good agreement with this result for sufficiently large difference in receding contact angles between the plates \citep{HCK16}. In the transition and dynamic regimes, and provided that inertia plays a secondary role, the transfer ratio, $\alpha$, can be estimated as \citep{CTA15b}:
\begin{equation}
\label{e2}
\alpha=0.5+\frac{\alpha_0-0.5}{1+p \text{Ca}^q},
\end{equation}
where $p$ and $q$ are fitting parameters. When inertia becomes important, $\alpha$ does not tend to 0.5 as the stretching speed increases.  Figure\ \ref{glycerol} shows the transfer of glycerol ($\rho=1.26$ kg/m$^3$, $\eta=1.42$ Pa$\cdot$s and $\gamma=63.4$ mN/m at 20$^{\circ}$) between Aluminium (AL) (donor surface) and poly(ethyl methacrylate) (PEMA) (acceptor surface). The receding contact angles for aluminium and PEMA are $37.5^{\circ}\pm 2.1^{\circ}$ and $57.3^{\circ}\pm 1.3^{\circ}$, respectively. As can be observed, the transfer ratio decreases as the velocity increases, and becomes $\sim$ 0.5 for $v\gtrsim 1$ mm/s ($\text{Ca}\gtrsim 0.0224$).

Despite its practical importance, the transfer of a non-Newtonian liquid between two flat surfaces with contact lines free to move is not fully understood. There are some studies in the literature for when contact line is pinned that have considered some rheological effects. Numerical simulations were carried out with the Carreau model to calculate the transfer ratio of a shear-thinning ink in gravure-offset printing \citep{GASS11} (in such systems contact lines are pinned on the surface). \citet{KBDR92} studied the influence of inertia, interfacial tension and elasticity on the stretching of a liquid bridge in a Plateau tank. \citet{SM96} measured the extensional viscosity of Boger fluids and found that this viscosity did not reach steady-state values for relatively small strain rates. \citet{YM98} analyzed the differences between the response of Newtonian and viscoelastic fluid filaments by solving the Oldroyd-B model. Their observations were in good qualitative agreement with experiments. \citet{SR12} studied experimentally the influence of viscoelasticity on the liquid transfer from an idealized gravure cell to a flat rigid substrate. They used a modified version of a capillary breakup rheometer in which the contact lines were pinned. The elastic stress significantly increased the bridge's lifetime and reduced the amount of fluid transferred from the gravure cell to the upper plate. Similar conclusions were obtained from the numerical simulations of the finitely extensible non-linear elastic (FENE-P) model \citep{LRP13}. In these simulations, elastic stresses were activated at early times for values of the Weissenberg number (which indicates the strain rate in terms of the polymer relaxation rate) exceeding a critical value. These stresses, combined with gravity, drain the liquid and reduce the volume transferred to the acceptor (upper) surface. \citet{KR17} studied experimentally the transfer of polyethylene oxide (PEO) in water from an idealized gravure cell to a rod. The liquid was subject to a combination of both shear and extensional deformations, and the contact line was pinned at the edge of the rod. The increase of the extensional viscosity enhanced the transfer. Overall, when the contact lines are pinned, the elastic stress sharply increases over the exponential thinning of the filament formed during the breakup of a liquid bridge. This increases the extensional viscosity, which stabilizes the liquid bridge and delays its breakup. The bridge eventually pinches at some point between the solid supports \citep{SR12,LRP13}. Viscoelasticity affects the mass transfer only quantitatively, i.e., $\alpha$ increases monotonically with the Capillary number, as occurs in the Newtonian case, but its value is affected by the elastic stress \citep{SR12}. 

As in the case of Newtonian liquids, the transfer of a non-Newtonian liquid between two solid surfaces can significantly change when the contact lines are unpinned. \citet{CK15} numerically solved the FENE-P model with moving contact lines to study the influence of viscoelasticity when a liquid is driven out of a cavity through both horizontal and vertical substrate motion. The results showed that viscoelasticity enhances the emptying of the cavity. \citet{WCK19} have recently studied the effects of both shear and extensional rheology on liquid transfer between two flat surfaces. Shear thinning enhances liquid transfer to the more-wettable surface because the viscosity reduction near the contact line on the less-wettable surface allows the contact line to slip more \citep{WCK18}. If surface-wettability difference is sufficiently large, then shear thinning allows nearly complete transfer for Capillary numbers (defined in terms of the zero-shear viscosity) greater than 0.1. Experiments of transfer of PEO between two flat surfaces show that a thin liquid thread is formed in the last stage of the process due to the growth of elastic stresses, which makes the contact radius decrease slowly during that stage \citep{WCK19}. Although the breakup time increases, the effect of elasticity on the amount of liquid transferred is relatively small. 

A natural question is how shear thinning and elasticity (extensional thickening) combined affect the liquid bridge evolution when the contact lines are allowed to move on the surfaces. Shear thinning favors the slip of the contact line on the surface, while elasticity delays the breakup of the liquid thread formed in the last phase of the process. One can expect that the combination of these two effects will produce the complete liquid transfer to the more-wettable surface even for very small Capillary numbers and surfaces with similar wettabilities. In this study, we will examine the above hypothesis, and discuss some practical implications.

\section{Methodology}
\label{sec2}

\begin{figure}
\begin{center}
\resizebox{0.5\hsize}{!}{\includegraphics{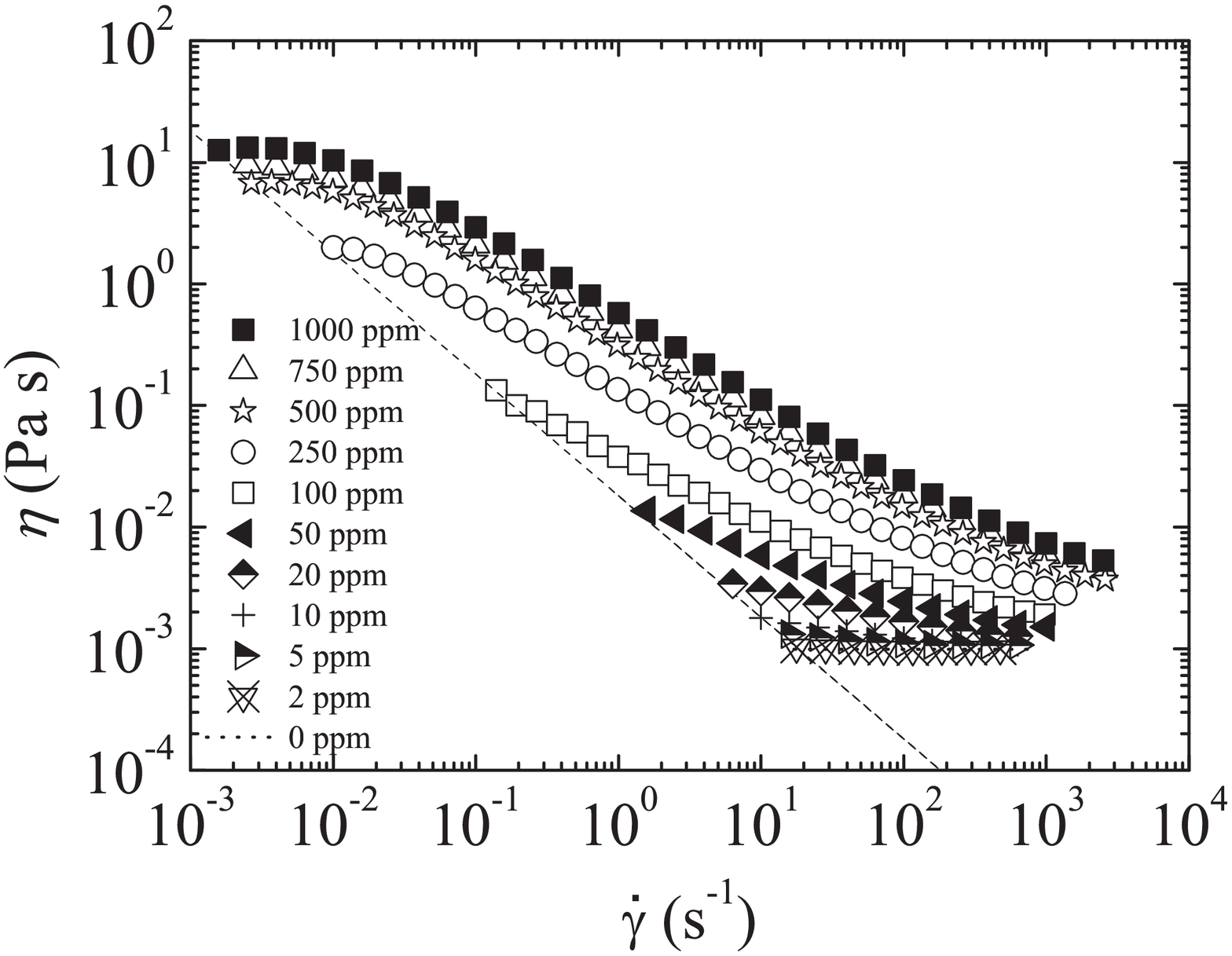}}\resizebox{0.45\hsize}{!}{\includegraphics{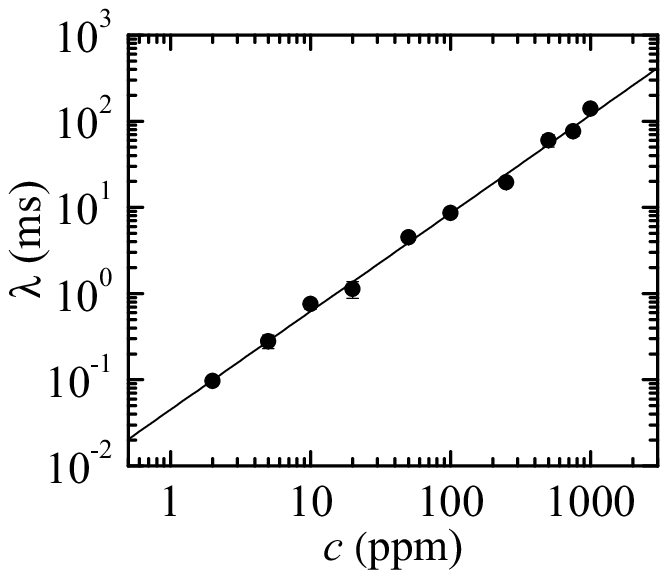}}
\end{center}
\caption{Shear viscosity $\eta$ as a function of the shear rate, $\dot{\gamma}$ (left panel), and extensional relaxation time $\lambda$ as a function of the polymer concentration, $c$ (right panel) both for PAA solutions at $T = 25~^{\circ}$C \citep{SVSMA17}. The dashed line in the left panel indicates the minimum measurable shear viscosity based on $20\times$ the minimum resolvable torque of the shear rheometer used. The line in the right panel represents the fit $\lambda$ [ms] $=0.045 (c$ [ppm])$^{1.14}$ to the data.}
\label{shear}
\end{figure}

Experiments were conducted with aqueous solutions of Poly(acrylic acid) (PAA) ({\sc Polysciences}) made at several concentrations ranging from 7.8 ppm to 1000 ppm. We chose this polymer because of its relatively large molecular weight ($M_w=18\times 10^6$ g/mol), which confers noticeable shear thinning and elasticity even at small concentrations. The PAA aqueous solution was carefully characterized by \citet{SVSMA17} for the range of concentrations considered in this work. The density and surface tension are about $\rho=1000$ kg/m$^3$ and $\gamma=70$ mN/m, respectively. The liquid exhibits significant shear thinning, while the extensional relaxation time, $\lambda$, depends almost linearly upon the polymer concentration, $c$ (Fig.\ \ref{shear}). Three surfaces were used in experiments: AL, silicon coated either with PEMA or polystyrene (PS). Details of the fabrication methods are given in the Supplementary Material. The receding contact angles between the polymeric solutions and these three surfaces were measured by the sessile drop method and found to be $\theta_r=39.1^{\circ}\pm 1.8^{\circ}$, $64.1^{\circ} \pm 2.0^{\circ}$ and $76.2^{\circ}\pm 2.2^{\circ}$ for AL, PEMA and PS, respectively. The receding contact angles are deemed important in liquid transfer studies \citep{CTA14}, therefore, the advancing contact angles were not needed. 

The acceptor surface was connected to a motion controller, while the donor surface was fixed to a horizontal platform. A droplet was gently placed on the donor surface. Due to the viscoelastic character of the solution, a small part of the liquid remained attached to the injection needle, which prevented the accurate control of the droplet volume. The droplet volume for the results presented in this paper was $\sim$ 1.6 $\mu$l. To ascertain whether or not small inaccuracies in volume will affect the transfer outcome, we also conducted experiments with volumes $\sim$ 1.2 $\mu$l (see next section and the Supplementary Material). The acceptor surface was moved slowly towards the donor surface until it touched the droplet and a liquid bridge was formed. The acceptor surface accelerated to a target speed ranging from 0.0125 to 15 mm/s in a very short time, and stayed at the set speed until the liquid bridge broke up or detached from one of the surfaces. Images of the stretching process were acquired with uniform backlight and a high-speed camera at 6\,700, 7\,400 or 40\,000 fps depending on the experiment. The images were processed with a sub-pixel resolution technique \citep{VMF11} to determine the position of the free surface as a function of time. The experiments were repeated at least three times to ensure the reproducibility of the results. Fresh surfaces were used in each experiment.

\section{Results and discussion}
\label{sec3}

\subsection{Effect of stretching speed}

\begin{figure}
\begin{center}
\resizebox{0.8\hsize}{!}{\includegraphics{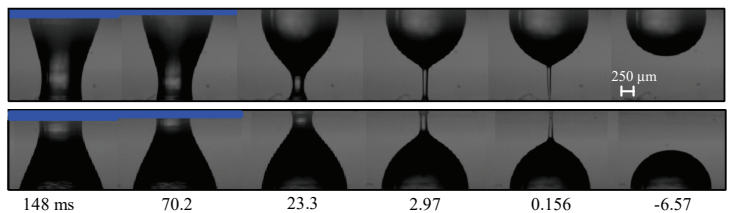}}
\end{center}
\caption{Transfer of water+PAA 1000 ppm from PS to PEMA (top row) and PEMA to PS (bottom row) for $v=1$ mm/s. The labels indicate the times to detachment (contact line fully receded). The scale bar applies to all the images. The initial distance between the surfaces was slightly different in the two experiments due to intrinsic contact angle differences between PS and PEMA. The liquid volume is $\sim$ 1.6 $\mu$l.}
\label{flip}
\end{figure}

\begin{figure}
\begin{center}
\resizebox{0.8\hsize}{!}{\includegraphics{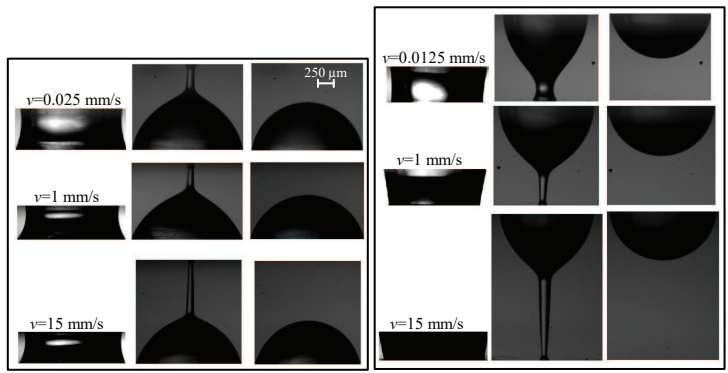}}
\end{center}
\caption{Transfer of water+PAA 1000 ppm from AL to PEMA (left) and from PS to PEMA (right). In each row, the sequence of images shows the intial liquid bridge shape, the shape in the last image before the full receding of the contact line, and the final state. The scale bar applies to all the images. The liquid volume is $\sim$ 1.6 $\mu$l.}
\label{altopema}
\end{figure}

We conducted experiments for $c=1000$ ppm and stretching speeds ranging from 0.0125 mm/s to 15 mm/s. In all cases, the contact line completely receded before the free surface pinched at its neck, resulting in detaching of the bridge from the surface with a higher receding contact angle. This means that the receding contact line was faster than the thinning of the bridge neck (see Fig.\ \ref{flip}). We verified that the observed behavior was independent of which surface of the two was donor or acceptor. For example, the liquid remained attached to the PEMA surface regardless of whether it was the donor or acceptor surface (Fig.\ \ref{flip}). In experiments with bridges between AL and PEMA, the liquid remained attached to the AL surface (Fig.\ \ref{altopema}-left). In all the experiments for liquid bridges between PS and PEMA, the liquid remained attached to the PEMA surface (Fig.\ \ref{altopema}-right). Finally, in all the experiments from PS to PS, the liquid always remained on the lower surface (see Supplementary Material). This behavior takes place regardless of the polymer concentration (provided that it exceeds a certain threshold), the liquid volume within the range 1.2-1.6 $\mu$l, and stretching speed. This constitutes a substantial difference with respect to what occurs to Newtonian liquids \citep{CTA14}. 

\subsection{Effect of contact angle}

For Newtonian liquids it was established that the difference between the receding contact angles of the acceptor and donor surfaces is the determining factor for the transfer ratio (for a given level of wettability of the donor and acceptor surfaces), see \citep{CTA15a,CTA15b,CTA14} and Eqs.\ (\ref{e1}) and (\ref{e2}). Furthermore, from Eqs.\ (\ref{e1}) and (\ref{e2}) it can be understood that for either of quasi-static or dynamic/transition regimes that change in transfer ratio (or how the liquid bridge will break up) is by in large a continuous function of wetting (i.e. the liquid transfer based on wetting changes from none to partial to complete transfer).  However, the discussion in Sec. 3.1 indicates that the entire liquid volume remains attached to the surface with the lowest receding contact angle for non-Newtonian fluid used here, no matter what is the concentration of polymer or speed of stretching. Hence the transfer ratio (a) is acting as a ``step function"\ since $\alpha=0$ if $\theta_r^{(\textin{acc})}\geq \theta_r^{(\textin{don})}$, and 1 otherwise. This finding is very different from what has been understood thus far based on studies of the bridge breakup and liquid transfer using Newtonian liquids; the same is true also for non-Newtonian fluids, but when contact line has been pinned on both or one of the surfaces bounding the liquid bridge, e.g. \citep{GASS11,KR17}. 

\subsection{Effect of polymer concentration}

To determine the threshold of the polymer concentration, $c$, for which this binary response takes places, we analyzed the liquid transfer from PS to PEMA for the stretching speeds $v=0.1$ and 10 mm/s, and $c=7.8$, 15.6, 31.25, 50, 62.5, 125, 250, 500 and 1000 ppm. The liquid was completely transferred to the PEMA surface for $c\geq 31.25$ ppm, while partial (Newtonian-like) transfer occurred for $c\leq 7.8$ ppm. In fact, $\alpha=0.98$ and 0.97 for $c=7.8$ ppm and $v=1$ and 10 mm/s, respectively. For the intermediate concentration $c=15.6$ ppm, the total or partial transfer took place, which is characteristic of a transitional behavior from the total to partial transfer regimes. The extensional relaxation time for this intermediate concentration is $\lambda=0.7$ ms (Fig.\ \ref{shear}). The fact that complete transfer took place even for relatively small values of $\lambda$ indicates that the combination of shear thinning and elasticity may produce the transition from partial to total transfer for a wide variety of polymeric solutions and concentrations.

\subsection{Role played by dimensionless numbers}

Total liquid transfer occured in our experiments for Capillary numbers in the range $5\times 10^{-6}$--2.7, which includes values of order unity typically found in high-speed printing applications. The Bond number, $B=\rho g {\cal V}^{2/3}/\gamma$, takes the values 0.16 and 0.19 for liquid bridges 1.2 and 1.6 $\mu$l in volume, respectively. These Bond number values are sufficiently large to play a critical role in the transfer of liquid between surfaces with the same wetting properties. 

Our experimental results are insensitive to the values taken by the Reynolds and Weber numbers because inertia plays a secondary role as compared to that of viscosity and surface tension. Consider, for instance, the experiment for c=1000 ppm and the largest stretching speed $v=15$ mm/s. The Reynolds number, Re$=\rho{\cal V}^{1/3}v/\eta_0$ defined in terms of the cube root of the bridge volume, ${\cal V}$, and the solution zero-shear viscosity, $\eta_0$, is $\sim 0.0014$. The strain rate in our experiments exceeded the critical value for the coil-stretch transition, and the extensional viscosity sharply increased. Therefore, the Reynolds number defined in terms of the extensional viscosity can be several orders of magnitude smaller than the above-mentioned value. 

The Weber number, We$=\rho v^2 {\cal V}^{1/3}/\gamma$, takes the value 0.0038 in the example considered above, which shows that this parameter is also irrelevant in our analysis. The Deborah number, De$=\lambda/t_c$, defined as the ratio of the stress relaxation time, $\lambda$, to the capillary time, $t_c=(\rho {\cal V}/\gamma)^{1/2}$, must play a relevant role in the experiment. This is implicitly assumed in establishing a critical value of the polymer concentration for the detaching of the bridge from the surface. However, the Deborah number alone cannot characterize the response of the system due the importance of the shear thinning in the liquid layer next to the solid surface. Our experiments show that the polymer concentration effectively accounts for both stress relaxation (quantified by the Deborah number) and shear thinning. This explains why there is a relatively well-established threshold of the polymer concentration above which total transfer of liquid takes place. 
 
 \subsection{Contact line dynamics}
 
\begin{figure}
  \begin{center}
  \includegraphics[width=0.275\linewidth]{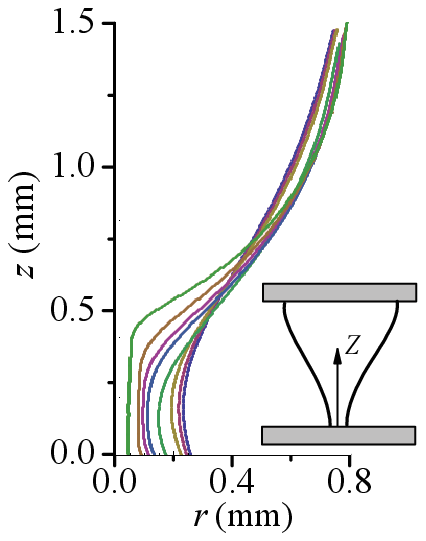}\includegraphics[width=0.475\linewidth]{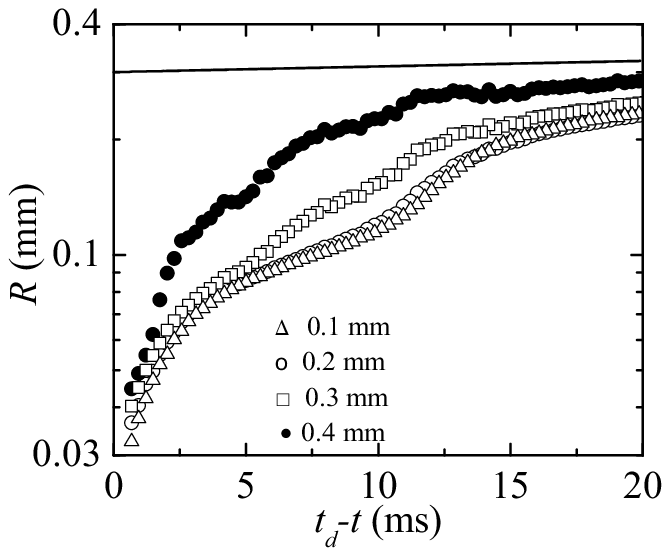}
  \end{center}
  \caption{(Left) Profile of the liquid bridge at times to detachment $t_d-t$=1.22, 3.92, 6.62, 9.32, 12.0, 14.7, 17.4 and 20.1 ms. (Right) Free surface radius $R(z)$ as a function of the time to detachment, $t_d-t$, for $z=0.1$, 0.2, 0.3 and 0.4 mm. The solid line corresponds to the exponential thinning $R=A \exp[(t_d-t)/(3 \lambda)]$ with $\lambda=100$ ms. The donor and acceptor surfaces were PS and PEMA, respectively. The acceptor surface velocity was 0.1 mm/s. The PAA concentration was 1000 ppm.}
  \label{contour}
\end{figure}

The dynamics of viscoelastic liquid bridges with moving contact lines are fundamentally different from those taking place in an extensional rheometer, in which the contact lines are pinned. As shown in Fig.\ \ref{contour}-left, the lower part of the liquid bridge narrows while the volume of the upper one increases. The contact line on the donor surface recedes while the contact line on the acceptor surface remains practically pinned. Figure \ref{contour}-right shows the free surface radius $R(z)$ as a function of the time to detachment $t_d-t$ for different distances, $z$, from the lower surface. The solid line corresponds to the exponential thinning taking place in the elastocapillary regime \citep{GYPS69} arising during the capillary breakup with pinned contact lines. As can be observed, the thinning of the filament formed above the donor surface is much faster than that exponential thinning. 

\begin{figure}
  \begin{center}
  \includegraphics[width=0.475\linewidth]{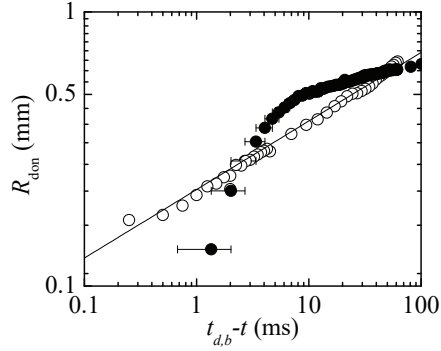}
  \end{center}
  \caption{Radius of the contact line on the donor surface, $R_{\textin{don}}$, as a function of the time to detachment/breakup $t_{d,b}-t$ for water (open symbols), and water+PAA 50 ppm (solid symbols). The solid line corresponds to the fit $R_{\textin{don}}=0.225 (t_{d,b}-t)^{1/4}$ to the water data. The donor and acceptor surfaces were PS and PEMA, respectively. The acceptor surface velocity was 10 and 1 mm/s for water and water+PAA 50 ppm, respectively.}
  \label{rlow}
\end{figure}

Figure \ref{rlow} compares the evolution of the contact line radius on the donor surface, $R_{\textin{don}}$, for water and water+PAA 50 ppm, both characterized by approximately the same Capillary number $\text{Ca}\simeq 10^{-4}$ (in the viscoelastic case, Ca is defined in terms of the highest measured value of $\eta(\dot{\gamma})$). In the Newtonian case, $R_{\textin{don}}$ approximately follows a power law until reaching a constant value, which occurs when the capillary breakup takes place. On the contrary, the contact line of the viscoelastic liquid bridge moves slower than that of its Newtonian counterpart until some instant shortly before the detachment ($t_d-t\sim 10$ ms), at which the contact line sharply accelerates. This instant likely corresponds to the growth of the elastic stress near the contact line due to the thinning of the filament above the surface. 

\begin{figure}
  \begin{center}
  \includegraphics[width=0.475\linewidth]{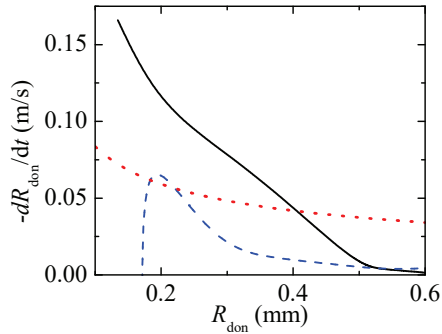}
  \end{center}
  \caption{Velocity of the contact line on the donor surface, $-dR_{\textin{don}}/dt$, as a function of the radius $R_{\textin{don}}$ for water (dashed line) and water+PAA 50 ppm (solid line). The dotted line is one tenth the inertio-capillary speed $V_{\textin{ic}}=(\rho R_{\textin{don}}/\gamma)^{-1/2}$. The donor and acceptor surfaces were PS and PEMA, respectively. The acceptor surface velocity was 10 and 1 mm/s for water and water+PAA 50 ppm, respectively.}
  \label{vlow}
\end{figure}

Figure \ref{vlow} shows the velocity $-dR_{\textin{don}}/dt$ of the contact line on the donor surface calculated from the smoothen data for $R_{\textin{don}}(t_{d,b}-t)$ shown in Fig.\ \ref{rlow}. In the Newtonian case, the contact line accelerates until its radius is approximately 0.2 mm, and then it sharply slows down until it reaches a static position. On the contrary, the contact line of the viscoelastic liquid bridge continuously accelerates until the bridge detaches from the donor surface. Both in the Newtonian and viscoelastic cases, the contact line speed is much smaller than the inertio-capillary speed $V_{\textin{ic}}=(\rho R_{\textin{don}}/\gamma)^{-1/2}$ characterizing the capillary breakup of a filament of radius $R_{\textin{don}}$ \citep{MP20}.

The results presented in this section can be explained in terms of the combined effects of shear thinning and elastic stress. Shear thinning enhances the slip of the contact line on the surface, while elasticity (extensional thickening) delays the breakup of the liquid thread formed (between surface and liquid bulk; see Fig.\ \ref{flip}) in the last stage of the process. As a result, the contact line fully retracts before the liquid bridge breakup takes place at its neck. This phenomenon occurs even for surfaces with very similar wettabilities and for Capillary numbers much smaller than those leading to the same phenomenon in the absence of elasticity \citep{WCK19}. For instance, the Capillary number corresponding to $c=1000$ ppm and $v=0.0125$ mm/s is $\sim 1.2\times 10^{-3}$, while the minimum value for complete contact line retraction in the absence of elasticity is O$(10^{-1})$ \citep{WCK19}.

\section{Concluding remarks}
\label{sec4}

We studied the effect of wettability, polymer concentration, and stretching speed on the breakup of a liquid bridge; the contact lines were free to move on the solid surfaces that delimit the bridge. We used aqueous solutions of PAA over three decades of concentrations; PAA solutions are interesting, as noticeable shear thinning and elasticity can be observed even at small concentrations (salient point being the density and surface tension remains fairly constant to remove their influence on the results of this study). We hypothesized that the competition between contact line dynamics and the rate of thinning of the bridge's neck is a dominant factor to determine the transfer ratio (bridge breakup). Our results showed that indeed the rate of receding of contact line is always higher than the rate of thinning of the liquid bridge's neck; this was not reported for either Newtonian liquids with free moving contact line, e.g. \citep{MP20,K15,CTA15a,CTA15b,CTA14}, or even for non-Newtonian liquids, with either free moving contact lines on both surfaces, e.g. \citep{WCK19}, or one of the surfaces, e.g. \citep{SR12}, or both pinned, e.g. \citep{GASS11,KR17}. The underlying mechanism is the delay of thinning of the bridge's neck due to the extensional thickening, and enhancement of contact line receding due to shear thinning.

We further showed regardless of the wettability, stretching speed (over three orders of magnitude), or liquid volume in the range studied, the liquid transfer will take a binary mode (i.e. either there will be a complete transfer or not at all). This behavior was not reported in any of the two recent comprehensive reviews of the topic \citep{MP20,K15}.  Again, shear thinning that favors facile slip of the contact line on the surface, combined with elasticity that slows the thinning of the bridge's neck in the last phase of the process is found responsible for our observation. One can expect that the combination of these two effects will produce the complete liquid transfer to the more-wettable surface even for very small Capillary numbers and for surfaces with similar wettabilities. We also showed that for polymer concentrations higher than a certain threshold ($\sim$ 30 ppm), the liquid volume remains attached to the surface with the lowest receding contact angle. For low concentrations, the shear thinning and elasticity of the liquid diminishes, and hence liquid transfer similar to that found in earlier studies, e.g. \citep{MP20,K15}, is seen.  For concentrations above 30 ppm, if the upper and lower receding contact angles are the same, gravity makes the liquid remain on the lower surface. These results are again independent of the range of liquid volume and stretching speed studied here. 

The results presented in this work have important implications in several technological applications; in particular, in offset printing with viscoelastic solutions where viscoelasticity may drastically change the liquid transfer. Complete liquid transfer can be achieved with polymeric solutions that exhibit both shear thinning and extensional thickening even for surfaces with very similar wettabilities and for very small stretching speeds. The use of polymeric solutions in capillary grippers \citep{ZMWA19} may entail significant advantages too. For instance, it can eliminate the liquid remaining on the released micro-object.

\section*{Acknowledgements}
 Partial support from the Ministerio de Econom\'{\i}a y Competitividad and Gobierno de Extremadura (Spain) through Grant Nos. DPI2016-78887 and GR18175 are gratefully acknowledged. AA acknowledges the funding from NSERC.

\section*{Appendix A: Supplementary material}
Supplementary material to this article can be found online at [the link to be inserted during production]. Details about the surfaces fabrication as well as additional results can be found in the Supplementary Material.

\newpage


\begin{thebibliography}{21}
\expandafter\ifx\csname natexlab\endcsname\relax\def\natexlab#1{#1}\fi
\providecommand{\bibinfo}[2]{#2}
\ifx\xfnm\relax \def\xfnm[#1]{\unskip,\space#1}\fi
\bibitem[{Montanero and Ponce-Torres(2020)}]{MP20}
\bibinfo{author}{J.~M. Montanero}, \bibinfo{author}{A.~Ponce-Torres},
\newblock \bibinfo{title}{Review on the dynamics of isothermal liquid bridges},
\newblock \bibinfo{journal}{Appl. Mech. Rev.} \bibinfo{volume}{72}
  (\bibinfo{year}{2020}) \bibinfo{pages}{010803}.
\bibitem[{Kumar(2015)}]{K15}
\bibinfo{author}{S.~Kumar},
\newblock \bibinfo{title}{Liquid transfer in printing processes: Liquid bridges
  with moving contact lines},
\newblock \bibinfo{journal}{Annu. Rev. Fluid Mech.} \bibinfo{volume}{47}
  (\bibinfo{year}{2015}) \bibinfo{pages}{67--94}.
\bibitem[{Snoeijer and Andreotti(2013)}]{SA13}
\bibinfo{author}{J.~H. Snoeijer}, \bibinfo{author}{B.~Andreotti},
\newblock \bibinfo{title}{Moving contact lines: Scales, regimes, and dynamical
  transitions},
\newblock \bibinfo{journal}{Annu. Rev. Fluid Mech.} \bibinfo{volume}{45}
  (\bibinfo{year}{2013}) \bibinfo{pages}{269--292}.
\bibitem[{Chen et~al.(2015)Chen, Tang, and Amirfazli}]{CTA15a}
\bibinfo{author}{H.~Chen}, \bibinfo{author}{T.~Tang},
  \bibinfo{author}{A.~Amirfazli},
\newblock \bibinfo{title}{Effects of surface wettability on fast liquid
  transfer},
\newblock \bibinfo{journal}{Phys. Fluids} \bibinfo{volume}{27}
  (\bibinfo{year}{2015}) \bibinfo{pages}{112102}.
\bibitem[{Huang et~al.(2016)Huang, Carvalho, and Kumar}]{HCK16}
\bibinfo{author}{C.~H. Huang}, \bibinfo{author}{M.~S. Carvalho},
  \bibinfo{author}{S.~Kumar},
\newblock \bibinfo{title}{Stretching liquid bridges with moving contact lines:
  comparison of liquid-transfer predictions and experiments},
\newblock \bibinfo{journal}{Soft Matter} \bibinfo{volume}{13}
  (\bibinfo{year}{2016}) \bibinfo{pages}{7457--7469}.
\bibitem[{Chen et~al.(2015)Chen, Tang, and Amirfazli}]{CTA15b}
\bibinfo{author}{H.~Chen}, \bibinfo{author}{T.~Tang},
  \bibinfo{author}{A.~Amirfazli},
\newblock \bibinfo{title}{Fast liquid transfer between surfaces: Breakup of
  stretched liquid bridges},
\newblock \bibinfo{journal}{Langmuir} \bibinfo{volume}{31}
  (\bibinfo{year}{2015}) \bibinfo{pages}{11470--11476}.
\bibitem[{Ghadiri et~al.(2011)Ghadiri, Ahmed, Sung, and Shirani}]{GASS11}
\bibinfo{author}{F.~Ghadiri}, \bibinfo{author}{D.~H. Ahmed},
  \bibinfo{author}{H.~J. Sung}, \bibinfo{author}{E.~Shirani},
\newblock \bibinfo{title}{{Non-Newtonian} ink transfer in gravure-offset
  printing},
\newblock \bibinfo{journal}{Int. J. Fluid Heat Flow} \bibinfo{volume}{32}
  (\bibinfo{year}{2011}) \bibinfo{pages}{308--317}.
\bibitem[{Kroger et~al.(1992)Kroger, Berg, Delgado, and Rath}]{KBDR92}
\bibinfo{author}{R.~Kroger}, \bibinfo{author}{S.~Berg},
  \bibinfo{author}{A.~Delgado}, \bibinfo{author}{H.~J. Rath},
\newblock \bibinfo{title}{Stretching behavior of large polymeric and
  {Newtonian} liquid bridges in {Plateau} simulation},
\newblock \bibinfo{journal}{J. Non-Newtonian Fluid Mech.} \bibinfo{volume}{45}
  (\bibinfo{year}{1992}) \bibinfo{pages}{385--400}.
\bibitem[{Solomon and Muller(1996)}]{SM96}
\bibinfo{author}{M.~J. Solomon}, \bibinfo{author}{S.~J. Muller},
\newblock \bibinfo{title}{The transient extensional behavior of
  polystyrene-based boger fluids of varying solvent quality and molecular
  weight},
\newblock \bibinfo{journal}{J. Rheology} \bibinfo{volume}{40}
  (\bibinfo{year}{1996}) \bibinfo{pages}{837--856}.
\bibitem[{Yao and McKinley(1998)}]{YM98}
\bibinfo{author}{M.~Yao}, \bibinfo{author}{G.~H. McKinley},
\newblock \bibinfo{title}{Numerical simulation of extensional deformations of
  viscoelastic liquid bridges in filament stretching devices},
\newblock \bibinfo{journal}{J. Non-Newtonian Fluid Mech.} \bibinfo{volume}{74}
  (\bibinfo{year}{1998}) \bibinfo{pages}{47-88}.
\bibitem[{Sankaran and Rothstein(2012)}]{SR12}
\bibinfo{author}{A.~K. Sankaran}, \bibinfo{author}{J.~P. Rothstein},
\newblock \bibinfo{title}{Effect of viscoelasticity on liquid transfer during
  gravure printing},
\newblock \bibinfo{journal}{J. Non-Newtonian Fluid Mech.} \bibinfo{volume}{175}
  (\bibinfo{year}{2012}) \bibinfo{pages}{64--75}.
\bibitem[{Lee et~al.(2013)Lee, Rothstein, and Pasquali}]{LRP13}
\bibinfo{author}{J.~A. Lee}, \bibinfo{author}{J.~P. Rothstein},
  \bibinfo{author}{M.~Pasquali},
\newblock \bibinfo{title}{Computational study of viscoelastic effects on liquid
  transfer during gravure printing},
\newblock \bibinfo{journal}{J. Non-Newtonian Fluid Mech.} \bibinfo{volume}{199}
  (\bibinfo{year}{2013}) \bibinfo{pages}{1--11}.
\bibitem[{Khandavalli and Rothstein(2017)}]{KR17}
\bibinfo{author}{S.~Khandavalli}, \bibinfo{author}{J.~P. Rothstein},
\newblock \bibinfo{title}{Ink transfer of {non-Newtonian} fluids from an
  idealized gravure cell: The effect of shear and extensional deformation},
\newblock \bibinfo{journal}{J. Non-Newtonian Fluid Mech.} \bibinfo{volume}{243}
  (\bibinfo{year}{2017}) \bibinfo{pages}{16--26}.
\bibitem[{Chung and Kumar(2015)}]{CK15}
\bibinfo{author}{C.~Chung}, \bibinfo{author}{S.~Kumar},
\newblock \bibinfo{title}{Emptying of viscoelastic liquids from model gravure
  cells},
\newblock \bibinfo{journal}{J. Non-Newtonian Fluid Mech.} \bibinfo{volume}{221}
  (\bibinfo{year}{2015}) \bibinfo{pages}{1--8}.
\bibitem[{Wu et~al.(2019)Wu, Carvalho, and Kumar}]{WCK19}
\bibinfo{author}{J.-T. Wu}, \bibinfo{author}{M.~S. Carvalho},
  \bibinfo{author}{S.~Kumar},
\newblock \bibinfo{title}{Effects of shear and extensional rheology on liquid
  transfer between two flat surfaces},
\newblock \bibinfo{journal}{J. Non-Newtonian Fluid Mech.} \bibinfo{volume}{274}
  (\bibinfo{year}{2019}) \bibinfo{pages}{104173}.
\bibitem[{Wu et~al.(2018)Wu, Carvalho, and Kumar}]{WCK18}
\bibinfo{author}{J.-T. Wu}, \bibinfo{author}{M.~S. Carvalho},
  \bibinfo{author}{S.~Kumar},
\newblock \bibinfo{title}{Transfer of rate-thinning and rate-thickening liquids
  between separating plates and cavities},
\newblock \bibinfo{journal}{J. Non-Newtonian Fluid Mech.} \bibinfo{volume}{255}
  (\bibinfo{year}{2018}) \bibinfo{pages}{57--69}.
\bibitem[{Sousa et~al.(2017)Sousa, Vega, Sousa, Montanero, and Alves}]{SVSMA17}
\bibinfo{author}{P.~C. Sousa}, \bibinfo{author}{E.~J. Vega},
  \bibinfo{author}{R.~G. Sousa}, \bibinfo{author}{J.~M. Montanero},
  \bibinfo{author}{M.~A. Alves},
\newblock \bibinfo{title}{Measurement of relaxation times in extensional flow
  of weakly viscoelastic polymer solutions},
\newblock \bibinfo{journal}{Rheol. Acta} \bibinfo{volume}{56}
  (\bibinfo{year}{2017}) \bibinfo{pages}{11--20}.
\bibitem[{Chen et~al.(2014)Chen, Tang, and Amirfazli}]{CTA14}
\bibinfo{author}{H.~Chen}, \bibinfo{author}{T.~Tang},
  \bibinfo{author}{A.~Amirfazli},
\newblock \bibinfo{title}{Mechanism of liquid transfer between two surfaces and
  the role of contact angles},
\newblock \bibinfo{journal}{Soft Matter} \bibinfo{volume}{10}
  (\bibinfo{year}{2014}) \bibinfo{pages}{2503--2507}.
\bibitem[{Vega et~al.(2011)Vega, Montanero, and Ferrera}]{VMF11}
\bibinfo{author}{E.~J. Vega}, \bibinfo{author}{J.~M. Montanero},
  \bibinfo{author}{C.~Ferrera},
\newblock \bibinfo{title}{Exploring the precision of backlight optical imaging
  in microfluidics close to the diffraction limit},
\newblock \bibinfo{journal}{Measurement} \bibinfo{volume}{44}
  (\bibinfo{year}{2011}) \bibinfo{pages}{1300--1311}.
\bibitem[{Goldin et~al.(1969)Goldin, Yerushalmi, Pfeffer, and Shinnar}]{GYPS69}
\bibinfo{author}{M.~Goldin}, \bibinfo{author}{J.~Yerushalmi},
  \bibinfo{author}{R.~Pfeffer}, \bibinfo{author}{R.~Shinnar},
\newblock \bibinfo{title}{Breakup of a laminar capillary jet of a viscoelastic
  fluid},
\newblock \bibinfo{journal}{J. Fluid Mech.} \bibinfo{volume}{38}
  (\bibinfo{year}{1969}) \bibinfo{pages}{689--711}.
\bibitem[{Zhang et~al.(2019)Zhang, Ma, Wang, and Aoyama}]{ZMWA19}
\bibinfo{author}{Q.~Zhang}, \bibinfo{author}{C.~Ma}, \bibinfo{author}{H.~Wang},
  \bibinfo{author}{H.~Aoyama},
\newblock \bibinfo{title}{A release method of micro-objects based on the liquid
  bridge configuration},
\newblock \bibinfo{journal}{J. Micromech. Microeng.} \bibinfo{volume}{29}
  (\bibinfo{year}{2019}) \bibinfo{pages}{045006}.

\end{thebibliography}

\end{document}